\documentstyle[12pt]{article}
\newcommand{\bq}{\begin{equation}}
\newcommand{\eq}{\end{equation}}
\newcommand{\bqa}{\begin{eqnarray}}
\newcommand{\eqa}{\end{eqnarray}}
\newcommand{\ra}{\rightarrow}

\def\half{{1 \over 2}}
\def\s{\sigma}

\def\D{\Delta}

\def\g{\gamma}

\def\ep{\epsilon}
\def\l{\lambda}

\def\ov{\over}

\def\rt{\sqrt{2}}
\def\ra{\rightarrow}

\def\2pi{1\over 2\pi i}
\def\q{q-q^{-1}}
\def\~{\tilde}
\def\newline{\hfil\break}

\def\ra{\rightarrow}

\def\sq2{\sqrt{2}}
\def\sqk2{\sqrt{2(k+2}}
\def\sqk{\sqrt{k}}

\def\be{\begin{equation}}
\def\ee{\end{equation}}
\def\br{\begin{array}}
\def\er{\end{array}}
\def\bea{\begin{eqnarray}}
\def\eea{\end{eqnarray}}
\def\nn{\nonumber}
\def\le{\lefteqn}
\def\curlra{\buildrel{\sim}\over\longrightarrow}
\def\qbinom#1#2{{#1}\atopwithdelims[]{#2}}
\newcommand{\uq}{U_q (su(2)_k)}

\def\vt{\tilde{\Phi}}
\def\vo{\phi}

\def\bE{{\bar E}^{-}}
\def\zE{E_0^{-}}
\def\oE{{\hat E}^{-}}
\def\cE{{\tilde E}^{-}}
\def\zP{{\phi}^0}
\def\oP{{\hat \phi}}
\def\cP{{\tilde \phi}}
\def\pl{\prod\limits}
\def\sl{\sum\limits}
\def\ap{a^{\prime}}
\def\bp{b^{\prime}}
\def\cp{c^{\prime}}
\def\ep{\epsilon}
\def\epp{\epsilon^{\prime}}
\def\no{\circ}
\begin{document}
\begin{titlepage}
\rightline{CRM-1896}
\rightline{hep-th/9307124}
\rightline{July 19, 1993}
\vbox{\vspace{12mm}}
\vspace{1.0truecm}
\begin{center}
{\LARGE \bf N-point Correlation Functions of the \\Spin-1 XXZ Model
}\\[12mm]
{\large A.H. BOUGOURZI$^{1}$  and  ROBERT A. WESTON$^{2}$}\\
[3mm]{\it Centre de Recherche Math\'ematiques,
Universit\'e de Montr\'eal\\
C.P. 6128-A, Montr\'eal (Qu\'ebec) H3C 3J7, Canada.}\\[15mm]
\end{center}
\begin{abstract}
\noindent We extend the recent approach of M. Jimbo, K. Miki, T. Miwa, and
A. Nakayashiki to
derive an integral formula for the N-point correlation functions
of arbitrary local
operators of the antiferromagnetic spin-1 XXZ model.
For this, we
realize the quantum affine symmetry algebra $U_q(su(2)_2)$ of level 2 and its
corresponding type I vertex operators in terms of a
deformed bosonic field free of
a background charge, and
a  deformed fermionic field. Up to GSO type projections, the Fock space is
already irreducible and therefore no BRST projections are involved.
This means that no screening charges with their Jackson
integrals are required.
Consequently, our N-point correlation functions are given
in terms of usual classical
integrals only, just as those  derived by Jimbo et al in the case of the
spin-1/2 XXZ model
 through the Frenkel-Jing bosonization of $U_q(su(2)_1)$.

\end{abstract}
\footnotetext[1]{
Email: {\tt bougourz@ere.umontreal.ca}
}
\footnotetext[2]{
Email: {\tt westonr@ere.umontreal.ca}
}

\end{titlepage}

\section{Introduction}

Solving many-body interacting quantum  (statistical-mechanical or
field-theoretic)
systems is still a
challenging problem even in low space-time dimensions.
Recently, in a remarkable paper, Davies et al. \cite{Daval92}  presented a new
method
for
solving the antiferromagnetic spin-1/2 XXZ model directly
in the thermodynamic limit. This contrasts with the
more conventional Bethe Ansatz approach, in which the
theory is  formulated on a finite chain before
taking the
thermodynamic limit in order to make the
Bethe equations more tractable. Davis et al's approach
is appealing
for two obvious reasons: firstly, because in addition to being able to
diagonalise the transfer matrix, it
is also possible to calculate the correlation functions, form factors,
eigenstates
and S-matrix of the model; and  secondly
because it highlights the role and makes extensive use
of the quantum affine symmetry of the model
which is present only in the thermodynamic limit. The  role of
this symmetry is
analogous to that of infinite-dimensional symmetries in conformal
field theory.

The Hamiltonian of the spin $1/2$ XXZ Heisenberg quantum
spin chain is  \cite{tafa79,KiRe87,Gaudin,Aff89}
\be {\cal H}_{XXZ}=-\half \sum_{i=-\infty}^{\infty} (\s_i^x \s_{i+1}^x
+\s_i^y \s_{i+1}^y + \Delta\s_i^z \s_{i+1}^z) \label{hamiltonian}, \ee
where $\D=(q+q^{-1})/2$ is an isotropy parameter.
Davis et al. consider this model in
the $\D<-1$ antiferromagnetic regime.
The Hamiltonian \ref{hamiltonian} acts formally on the
infinite tensor product,
\be \cdots V \otimes V \otimes V \otimes V \cdots \label{infprod}\eq
where $V$ is a two dimensional representation of $U^\prime_q(su(2)_1)$  (in
what follows the subscript $k$ in $\uq$ or $U^\prime_q
(su(2)_k)$ stands for the
level of these algebras).
The latter is the symmetry algebra of \ref{hamiltonian} and is a subalgebra
(by dropping the grading operator) of the full quantum affine algebra
$U_q(su(2)_1)$.
  The central idea of the approach
of Davies et al.  is to replace this infinite tensor product by the
 $U_q(su(2)_0)$ module
\be{\cal F}_{\lambda,\mu} ={\rm Hom}(V(\mu),V(\lambda)),\ee
where $V(\lambda)$ and $\lambda=\Lambda_0,\Lambda_1$ are the $U_q(su(2)_1)$
  highest weight modules and highest weights respectively.
Following Frenkel and Reshetikhin \cite{FrRe92}, Davies et al. realised this
homomorphism  by introducing the following  vertex
operators (VOs) that intertwine the $U_q(su(2)_1)$
modules
\bq \vt^{\mu, V}_{\lambda}(z):V(\lambda) \ra V(\mu) \otimes V(z).
\eq
Here $z$ is a spectral parameter
and $V^j(z)$ is
the  `evaluation representation' of $U_q(su(2)_0)$, and
 is isomorphic to
$V\otimes {\bf C}[z,z^{-1}]$. These VOs 1.4  are the
type I VOs of
ref. \cite{Daval92}. By introducing a variant of these type I VOs, i.e.,
\bq \vt^{\mu}_{\lambda ,V}(z): V(\lambda) \otimes V(z) \ra V(\mu),
\eq
Jimbo et al. \cite{collin}  were able to re-formulate the action of local
operators of the
vertex model (such as the spin local operator $\s_i^z$)
on the infinite product \ref{infprod}
in terms of their action on ${\cal F}_{\lambda,\mu}$.
In this manner, and by bosonizing the VOs, they were
able to write down an integral formula for N-point correlation functions
of local operators in the 6 vertex model.
Evaluating this integral for N=1, enabled them to triumphantly reproduce
Baxter's result for the staggered polarization of the 6-vertex
model \cite{bax82}.

Generalisations of the 6-vertex model, corresponding to spin-$S$ quantum
chains,
were introduced by Zamolodchikov
and Fatteev \cite{ZaFa80} and Kulish and
Reshitikhin \cite{KuRe81}.
The theory associated with $S=1$ is a 19-vertex model \cite{ZaFa80,KuRe81}.
The Hamiltonian is that of the antiferromagnetic XXZ
model \cite{ZaFa80,KiRe87,AlMa89}
\bea
\le{ {\cal H}=J\sl_{i=-\infty}^{\infty} \left\{S_i\cdot S_{i+1} -(S_i\cdot
S_{i+1})^2
+\half (q-q^{-1})^2
[S_i^z\cdot S_{i+1}^z -(S_i^z\cdot S_{i+1}^z)^2+2 (S_i^z)^2]\right.}\\
&&\left. -(q+q^{-1} -2)
[(S_i^x\cdot S_{i+1}^x +S_i^y\cdot S_{i+1}^y)S_i^z\cdot S_{i+1}^z
+S_i^z\cdot S_{i+1}^z(S_i^x\cdot S_{i+1}^x +S_i^y\cdot S_{i+1}^y)]\right\},
\nn\label{1hamil}\eea
where $S^x,S^y$ and $S^z$ are $(3\times 3)$ spin-1 matrices,
and $J>0$. This  Hamiltonian is symmetric under
$U_q^\prime(su(2)_2)$ \cite{idzal93}.
The theory is massless for $q=\exp(i \theta), ~ 0\leq
\theta\leq
\pi$ \cite{Sog84,what?}, and for  $q=1$ it reduces to the XXX
model discussed for example in  \cite{Aff89,Res91}.
We consider it instead in the massive
 region with
$q$ real and $-1<q<1$.
The general spin-$S$ Hamiltonian is a complicated
polynomial in the $2S+1$ dimensional spin matrices, and is given in the
paper by Sogo \cite{Sog84}.

Idzumi et al. \cite{idzal93} have extended the work of Davis et al. to
general level $k$.  The associated spin/vertex models are the generalised
XXZ models with $S=k/2$ .
The step of bosonizing the currents and vertex operators is however more
complicated for $k>1$. Firstly, one conventionally uses a deformation of the
Wakimoto bosonization \cite{BoGr92,Abaal92,Mat921,Mat922,Shi92,Katal92,Kim92}
- which requires three bosonic
fields unlike the Frenkel-Jing
bosonization for the case $k=1$ \cite{FrJi88}, which requires only one.
Secondly, and more seriously
when calculating correlation functions it is necessary to take a trace over
a $U_q(su(2)_k)$  irreducible highest weight module. For $k>1$, the bosonic
Fock space can no longer
be equated with any $U_q(su(2)_k)$  irreducible highest weight module,
and one must therefore take its
non-trivial BRST cohomology structure into account. This  BRST structure
can be studied by means of screening charges \cite{Fel90}, but the latter
are given in terms of Jackson integrals, and a formula for the N-point
correlation function derived using this approach
would be highly complicated.
In this paper, we realise the  $U_q(su(2)_2)$  algebra and its
type I vertex operators  in terms of one boson and one fermion. The major
advantage of this realisation is that now, up to a GSO-like projection
(though not exactly the familiar one),
the Fock space is irreducible as for $k=1$. Exploiting this simplicity, we
are able to produce an integral formula for the N-point correlation function
of local operators of the spin-1 XXZ chain.

The paper is organized as follows. In section 2, we fix the notation and
make this paper self contained
by reviewing the basic properties
of the $\uq$ quantum affine algebra. For the purpose of
free field realization we write this algebra  in an operator
product expansion form that we refer to as the
$\uq$ quantum current algebra.
In section 3, we use this quantum current algebra to realize
$U_q(su(2)_2)$  and its associated type I
vertex operators in terms of one boson and one fermion. Then we define the
Fock space built from the modes of these two fields, and describe how, with
the aid of a GSO-like projector, it is related to the
highest weight representations of $U_q(su(2)_2)$.
In section 4, we describe how the local operators of the spin-1
XXZ model act in ${\cal F}$, the space of excitation over the ground state.
Here we introduce the variants of type I vertex operators.  In section 5,
we use the realization of section 3 to write an
integral formula for the N-point correlation functions of arbitrary local
operators. All integrals involved here are the usual classical ones
and not Jackson integrals. Finally, section 6 is devoted to our conclusions.

\section { The $U_q(su(2)_k)$ Quantum Current Algebra}
In this section we review some basic properties of the $U_q(su(2)_k)$ quantum
affine algebra at level $k$, which will make the subsequent sections more
transparent. This algebra reads in the
Chevalley basis as \cite{Jim85,Dri85,Dri88}
\be\br{rcl}
&&t_it_j=t_jt_i,\\
&&t_i e_i t^{-1}_i=q^2e_i,\quad t_i e_j t^{-1}_i=q^{-2}e_j, ~~i\neq j,\\
&&t_i f_i t^{-1}_i=q^{-2}f_i,\quad t_i f_j t^{-1}_i=q^2f_j, ~~i\neq j,\\
&&[e_i, f_j]=\delta_{i,j}{t_i-t_i^{-1}\over q-q^{-1}},\\
&&q^de_iq^{-d}=q^{\delta_{i,0}}e_i,\quad q^df_iq^{-d}=q^{-\delta_{i,0}}f_i\quad
q^dt_iq^{-d}=t_i,
\label{cheval}
\er\ee
where $\{e_i,f_i,t_i,\,\, i,j=0,1\}$
are  the usual Chevalley generators and $d$ is the grading operator.
Its main feature is that it is an associative Hopf algebra with
  comultiplication
\be\br{rcl}
\Delta(e_i)&=&e_i\otimes 1+t_i\otimes e_i,\\
\Delta(f_i)&=&f_i\otimes t_i^{-1} +1\otimes f_i, \\
\Delta(t_i)&=&t_i\otimes t_i,\\
\Delta(q^d)&=&q^d\otimes q^d,
\label{comul}\er\ee
and antipode
\be
a(e_i)=-t^{-1}_ie_i,\qquad a(f_i)=-f_it_i,\qquad a(t_i)=t_i^{-1},\qquad
a(q^d)=q^{-d},\qquad
i=0,1.
\label{antip}
\ee
When  $d$ is omitted, the above algebra is referred to as
$U_q^\prime(su(2)_k)$. For many practical purposes, it is convenient to use the
Drinfeld realization of \ref{cheval} \cite{Dri85}. This is constructed from the
following
redefinitions of the Chevalley generators $\{e_i,f_i,t_i\}$:
\be\br{rcl}
t_0&=\gamma^k q^{-\rt H_0},\quad t_1&=q^{\rt H_0},\\
e_0&=E^-_1q^{-\rt H_0},\quad e_1&=E^+_0,\\
f_0&=q^{\rt H_0}E^+_{-1},\quad f_1&=E^-_0.
\label{ident}
\er\ee
Using  \ref{ident} and \ref{cheval}, one recursively generates the following
algebra, which is known as the Drinfeld realization of $U_q(su(2)_k)$:
\be\br{rcl}
&&{[H_n,H_m]} = {[2n]\over 2n} {{\g^{nk}-\g^{-nk}} \ov {q-q^{-1}}}
\delta_{n+m,0},\qquad
n\neq 0,\\
&&{[q^{\pm\rt H_0}, H_m]}=  0,\\ && {[H_n,E^{\pm}_m]}=
\pm\sqrt{2}{\g^{\mp |n|k/2}[2n]\over 2n}
E^\pm_{n+m}, \qquad n\neq 0,\\
&&  q^{\rt H_0}  E^\pm_n q^{-\rt H_0} =q^{\pm 2} E^\pm_n,\\
&&{[E^+_n,E^-_m]} = {\g^{k(n-m)/2}\psi_{n+m}-\g^{k(m-n)/2}
\varphi_{n+m}\over q-q^{-1}},\\
&&E^\pm_{n+1}E^\pm_m-q^{\pm 2}E^\pm_mE^\pm_{n+1}=
q^{\pm 2}E^\pm
_nE^\pm_{m+1}-E^\pm_{m+1}E^\pm_n,\\
&&q^dE^\pm_nq^{-d}=q^nE^\pm_n,\quad q^dH_nq^{-d}=q^nH_n.
\label{cwb}
\er\ee
This is an algebra generated by the Drinfeld operators
$\{ E^{\pm}_n~(n\in {\bf Z}),H_m ~(m \in {\bf Z_{ \neq 0}}),q^{\pm \rt H_0},
q^{\pm d}, \g^{\pm 1/2}\}$, where $\g^{\pm 1/2}$ commute with all the
generators,
and  as usual $[x]=(q^x-q^{-x})/(q-q^{-1})$. $\psi_n$ and $\varphi_n$ are
the modes of fields
$\psi(z)$ and $\varphi(z)$ defined by
\be\br{rcl}
\psi(z)&=&\sum\limits_{n\geq 0}\psi_nz^{-n}=q^{\sqrt{2}H_0}
 \exp\{\sqrt{2}(\q)\sum\limits_{n>0}H_nz^{-n}\},\\
\varphi(z)&=&\sum\limits_{n\leq 0}\varphi_nz^{-n}=q^{-\sqrt{2}H_0}
\exp\{-\sqrt{2}(\q)\sum\limits_{n<0}H_nz^{-n}\}.
\label{algebra}\er\ee
Using the basic identifications \ref{ident} one can re-express the
comultiplication \ref{comul} in terms of the Drinfeld generators \ref{cwb} as
\be\br{rcl}
\!\!\!\!\!\!& &\Delta(E^+_n)=E^+_n\otimes\gamma^{kn}+
\gamma^{2kn}q^{\sq2 H_0}\otimes
E^+_n+ \sum_{i=0}^{n-1}\gamma^{k(n+3i)/2}\psi_{n-i}\otimes \gamma^{k(n-i)}
 E^+_i\: {\rm mod}   \: {N_-}\otimes {N_+^2},\\
 \!\!\!\!\!\!& &\Delta(E^+_{-m})=E^+_{-m}\!\otimes\!\gamma^{-km}\!+\!
q^{-\sq2 H_0}\!\otimes\! E^+_{-m}+
\sum_{i=0}^{m-1}\gamma^{{k(m-i)\over 2}}\varphi_{-m+i}\otimes \gamma^{k(i-m)}
 E^+_{-i}
\: {\rm mod  }\:
 N_-\otimes N_+^2,\\
\!\!\!\!\!\!& &\Delta(E^-_{-n})=E^-_{-n}\!\otimes\!\gamma^{-2kn}q^{-\sq2
H_0}\!+\!
  \gamma^{-kn}\!\otimes \!E^-_{-n}\!+\!\!\sum_{i=0}^{n-1}\!\gamma^{-k(n-i)}
  E^-_i\!\otimes\!\gamma^{{-k(n+3i)\over 2}}\varphi_{i-n}\: {\rm mod  }\:
 N_-^2\!\otimes\! N_+,\\
\!\!\!\!\!\!& &\Delta(E^-_m)=\gamma^{km}\otimes E^-_m+E^-_m\otimes
q^{\sq2 H_0} +
\sum_{i=1}^{m-1}\gamma^{k(m-1)}E^-_m\otimes \gamma^{-k(m-i)/2}
\psi_{m-i}\:{\rm mod  }\: N_-^2\otimes N_+,\\
\!\!\!\!\!\!& &\Delta(H_m)=H_m\otimes\gamma^{km/2}+
\gamma^{3km/2}\otimes H_m\: {\rm mod  }\: N_-\otimes N_+,\quad\\
\!\!\!\!\!\!& &\Delta(H_{-m})=H_{-m}\otimes\gamma^{-3km/2}+\gamma^{-km/2}
\otimes  H_{-m}\: {\rm mod  }\: N_-\otimes N_+,\quad\\
\!\!\!\!\!\!& &\Delta(q^{\pm \sq2 H_0})=q^{\pm \sq2 H_0}\otimes
 q^{\pm \sq2 H_0},\quad\\
\!\!\!\!\!\!& &\Delta(\gamma^{\pm \half})=\gamma^{\pm\half}\otimes
\gamma^{\pm \half},\\
\!\!\!\!\!\!& &\Delta(q^{\pm d})=q^{\pm d}\otimes q^{\pm d},
\label{comult}\er\ee
where $m>0$, $n\geq 0$, and $N_\pm$ and $N_\pm^2$ are
left ${\bf Q}(q)[\gamma^\pm,  \psi_m, \varphi_{-n}; \: m, n\in {\bf
Z_{\geq 0}}]$ modules
generated
 by $\{E^\pm_m; \:m\in {\bf Z}\}$
and $\{E^\pm_m E^\pm_n; \:m, n\in {\bf Z}\}$  respectively
 \cite{collin,ChPr91}.
The main virtue of this comultiplication is that it will
make the derivation of the intertwining
properties of the
vertex operators explicit. These intertwining relations in turn allow the
free field
realizations (in terms of either free bosons, fermions, or ghosts) of the
vertex operators.
However for the purpose of the free field realization,
it is convenient to rewrite the algebra \ref{cwb}
 in terms of operator product expansions (OPEs), that is,  as
a quantum current algebra (QCA).
The $U_q(su(2)_k)$ QCA
then reads   \cite{Abaal92,Mat921,Mat922,Ber89}
\be\br{rcl}
\psi(z).\varphi(w)&=&
{(z-wq^{2+k})(z-wq^{-2-k})\over (z-wq^{2-k})(z-wq^{-2+k})}
\varphi(w).\psi(z), \\
\psi(z).E^{\pm}(w)&=&
q^{\pm 2}{(z-wq^{\mp(2+k/2)})\over z-wq^{\pm (2-k/2)}}
E^\pm(w).\psi(z), \\
\varphi(z).E^{\pm}(w)&=&
q^{\pm 2}{(z-wq^{\mp(2-k/2)})\over z-wq^{\pm (2+k/2)}}
E^\pm(w).\varphi(z),\\
E^+(z).E^-(w)&\sim& {1
\over w(\q)}\left\{{\psi(wq^{k/2})\over z-wq^k}-
{\varphi(wq^{-k/2})\over z-wq^{-k}}\right\},\quad |z|>|wq^{\pm k}|,\\
E^{\pm}(z).E^{\pm}(w)&=&{(z q^{\pm 2}-w)\over z-w q^{\pm 2}}
E^{\pm}(w). E^{\pm}(z).
\label{ope8}
\er\ee
Here the quantum currents $\psi(z)$ and $\varphi(z)$ are given
by \ref{algebra}, whereas $E^\pm(z)$ are the following generating functions in
terms of the Drinfeld generators:
\be
E^\pm(z)=\sum_{n=-\infty}^{+\infty}E^\pm_nz^{-n-1}.
\ee

The free field realization of the $U_q(su(2)_k)$ QCA therefore translates into
solving the relations \ref{ope8} in terms of free fields.
For example, the bosonization (i.e., the free field realization in terms
of bosons) of $\uq$ has recently been studied intensively. For $k=1$ this is
 known as the Frenkel-Jing bosonization and requires only a single deformed
boson field  \cite{FrJi88}. For general $k$, there
are many bosonizations available in the literature. We shall
refer to these as  q-deformations
of the Wakimoto bosonization
 \cite{BoGr92,Abaal92,Mat921,Mat922,Shi92,Katal92,Kim92,bou93}.
(See  \cite{bou93} for a detailed
discussion of the equivalence of these different
bosonizations.) Very recently, the bosonization of $U_q(su(n)_k)$ has been
constructed in Ref. \cite{sln}, and the difference
realizations of  $U_q(su(n))$ and general quantum Lie
algebras have been achieved in Refs.  \cite{other} and  \cite{vinet}
respectively.

\subsection{Intertwining relations of the vertex operators }
The vertex operators relevant to this discussion are
the type I intertwiners  of Refs  \cite{Daval92,idzal93}.
They are defined as maps between $\uq$ modules  in
the following way:
\bq \vt^{\mu,V^j}_{\lambda}(z):V(\lambda) \ra V(\mu) \otimes V^j(z).
\label{origvo}\eq
Here $V(\lambda)$ are $\uq$ left highest weight
modules, with $\{\lambda=\lambda_{i}=(k-i)\Lambda_0+i\Lambda_1,~ i=0,\dots,k\}$
and $\{\Lambda_0,\Lambda_1\}$ denoting the sets of $\uq$ dominant highest
weights and fundamental weights respectively. $V^j(z)$ $(0\leq j\leq k/2)$
is the spin $j$
 `evaluation representation' of $U_q(su(2)_0)$.
It is isomorphic to $V^j\otimes {\bf C}[z,z^{-1}]$, where $V^j$ is the
$U_q^\prime(su(2)_0)$ $2j+1$ dimensional representation,
 with the basis $\{v_m^j,~ -j\leq m\leq j\}$
such that:
\be\br{rcl}
e_1v^j_m=[j+m]v^j_{m-1},\quad f_1v^j_m=[j-m]v^j_{m+1},\quad
t_1v^j_m=q^{-2m}v^j_m,\\
e_0v^j_m=[j-m]v^j_{m+1},\quad f_0v^j_m=[j+m]v^j_{m-1},\quad
t_0v^j_m=q^{2m}v^j_m,
\er\ee
where it is understood that $v^j_m=0$ if $|m|>j$.
$V^j(z)$ is equipped with the following $U_q^\prime(su(2)_0)$ module
structure  \cite{idzal93}:
\be\br{rcl}
e_1v^j_m\otimes z^n&=&[j+m]v^j_{m-1}\otimes z^n,\\
e_0v^j_m\otimes z^n&=&[j-m]v^j_{m+1}\otimes z^{n+1},\\
f_1v^j_m\otimes z^n&=&[j-m]v^j_{m+1}\otimes z^n,\\
f_0v^j_m\otimes z^n&=&[j+m]v^j_{m-1}\otimes z^{n-1},\\
t_1v^j_m\otimes z^n&=&q^{-2m}v^j_m\otimes z^n,\\
t_0v^j_m\otimes z^n&=&q^{2m}v^j_m\otimes z^n.
\er\ee
In terms of the Drinfeld realization this becomes
\be\br{rcl}
\gamma^{\pm 1/2}v^j_m\otimes z^\ell&=&v^j_m\otimes z^\ell,\\
q^{\rt H_0} v_m^j\otimes z^\ell&=&q^{-2m}v^j_m\otimes z^\ell,\\
E^+_nv^j_m\otimes z^\ell&=&q^{2n(1-m)}[j+m]v^j_{m-1}\otimes z^{\ell+n},\\
E^-_nv^j_m\otimes z^\ell&=&q^{-2nm}[j-m]v^j_{m+1}\otimes z^{\ell+n},\\
H_nv^j_m\otimes z^\ell&=&{1\over{\rt n}}\{[2nj]-q^{n(j-m+1)}(q^n+q^{-n})
[n(j+m)]\}v^j_m\otimes z^{\ell+n}.
\er\ee
Let $V^{j*}_z$ be the evaluation representation dual to $V^j_z$ and endowed
with
 the left $U_q^\prime(su(2)_0)$  module structure through the action of the
antipode \ref{antip},
which is an
anti-automorphism of this algebra, i.e.,
\be
<xu,v>=<u,a(x)v>,\quad x\in U_q(su(2)_0),\qquad u\in V^{j*}_z,\qquad v\in
V^j_z.
\ee
Then it can easily be shown that the following two evaluation representations
are isomorphic to each other \cite{idzal93}:
\be
C:\, V^j_{zq^{-2}}\curlra V^{j*}_z,
\ee
where
\be\br{rcl}
Cv^j_m\otimes (zq^{-2})^n&=&{\cal C}_m^{j}{v^{j*}_{-m}}\otimes z^n
,\\
{\cal C}_m^j&=&(-1)^{j+m}q^{-(j+m)(j-m-1)}{\qbinom{2j}{j+m}}^{-1}
,\quad -j\leq m\leq j. \er\ee

Here $\{v^{j*}_m,\, -j\leq m\leq j\}$ is the basis of $V^{j*}$, which is dual
to $V^j$, and  the notation
$\qbinom{x}{y}$ defines the q-analogue of the
binomial coefficient as
\be\br{rcl}
\qbinom{x}{y}&=&{[x]!\over [y]![x-y]!},\\
{[x]}!&=&[x][x-1]\dots [1].
\er\ee

By definition the vertex operators $\vt^{\mu, V^j}_\lambda(z)$
obey the intertwining condition \cite{FrRe92,Daval92,idzal93}
\bq \vt^{\mu, V^j}_\lambda(z) \circ x = \Delta(x)\circ
\vt^{\mu, V^j}_\lambda(z) ~~~~~\forall~ x\in \uq .
\label{int} \ee
Here $x$ denotes a  generator in the Drinfeld realization,
which is appropriate for
explicitly constructing the vertex operators in terms of free fields.
It is also convenient to define components of these vertex operators through
\be \vt^{\mu, V^j}_\lambda(z) = s^{\mu, V^j}_{\lambda}(z)
\sum\limits_{m=-j}^{j}  \vo^j_m(z)\otimes v^{j}_m ,\ee
where the normalisation
function $s^{\mu, V^j}_{\lambda}(z)$ is given by \cite{BoWe93a}
\be s^{\mu, V^j}_{\lambda}(z)=(-zq^{k+2})^{(\D(\lambda_{2j})+
\D(\lambda)-\D(\mu))},\ee
with
\be
\D(\lambda_{2j})={j(j+1)\over k+2}.
\ee
$\vo^j_m(z)$ are referred to as the bare vertex operators.
Using the above relation \ref{int}, the
comultiplication \ref{comult}, and the fact that
$N_+v^j_{-j}=N_-v^j_j=0$, $N_{\pm}v^j_m\in
F[z,z^{-1}]
v^j_{m\mp 1}$, we arrive at the following commutation relations:
\be\br{rcl}
{[E^{+}(w),\vo^j_j(z)]}&=&0,\\
{[H_n, \vo^j_j(z)]}&=&j\rt \left\{\delta_{n,0}+
q^{(n(k+2)+|n|k/2)} { {[2jn]}\ov {2jn} }z^n \right\}\vo^j_j(z),\\
\vo^j_m(z)&=&{1\over {[j-m]!}}[\dots[\vo^j_j(z),E^{-}_0]_{q^{2j}}\dots,
E^-_0]_{q^{2(m+1)}}.
\label{phijm}\er\ee
In \ref{phijm} there are $(j-m)$ quantum commutators,
where the quantum commutator $[A,B]_{q^{x}}$ is defined by
\be\br{rcl}
[A,B]_{q^x}=AB-q^xBA.
\er\ee

\section{Realization of $U_q(su(2)_2)$ in Terms of One Boson and One Fermion}

When $k=2$,
one needs one  boson, deformed in two different ways as$\chi^\pm(z)$,
and one deformed fermion
$\Psi(z)$ in order to realize the quantum current
algebra \ref{ope8}. The quantum currents are given in terms of these
fields as
\be\br{rcl}
\psi(z)&= &\exp
\left(i\chi^{+}(zq)-i\chi^{-}(zq^{-1})\right)\\
&=&q^{2a_0}
\exp\left(2(\q)\sl_{n>0}a_nz^{-n}\right),\\
\varphi(z)&= &\exp
\left(i\chi^{+}(zq^{-1})-i\chi^{-}(zq)\right)\\
&=&q^{-2a_0}
\exp\left(-2(\q)\sl_{n<0}a_nz^{-n}\right),\\
E^\pm(z)&=&\sqrt{2}\Psi(z)\exp\left(\pm i\chi^{\pm}(z)\right),
\label{reali}
\er\ee
where
\be\br{rcl}
\chi^\pm(z)&=&a-ia_0\ln{z}
+2i \sl_{n>0}
{ q^{\mp n}a_n z^{-n}\over [2n]}
+2i \sl_{n<0}
{ q^{\pm n}a_n z^{-n}\over [2n]},\\
\Psi(z)&=&\sl_{r\in {\bf Z}+1/2}b_rz^{-r -1/2}.
\er\ee
The bosonic modes $\{a,a_n\}$ of $\chi^\pm(z)$  and the fermionic modes $b_r$
in the
Neuveu Schwartz sector of $\Psi(z)$ satisfy the following
Heisenberg algebra and anticommutation relations respectively:
\be\br{rcl}
{[a_n,a_m]}&=&{[2n]^2\over 4n}\delta_{n+m,0},\\
{[a,a_0]}&=&i,\\
\{b_r,b_s\}&=&{[4r]\over 2[2r]}\delta_{r+s,0}.
\er\ee
All the other commutation and anticommutation relations are
trivial.  $\chi^\pm(z)$ and $\Psi(z)$ have the
following simple OPEs:
\be\br{rcl}
\chi^\pm(z).\chi^\mp(w)&=&-\ln(z-w)+:\chi^\pm(z).\chi^\mp(w):,\quad |z|>|w|,\\
\chi^\pm(z).\chi^\pm(w)&=&-\ln(z-wq^{\mp 2})+
:\chi^\pm(z).\chi^\pm(w):,\quad |z|>|q^{\pm 2}w|,\\
\Psi(z).\Psi(w)&=& {[2]\over 2}\,{z-w\over (z-wq^2)(z-wq^{-2})}+
:\Psi(z)\Psi(w):,\quad |z|>|q^{\pm 2}w|.
\label{ferm}
\er\ee
It is clear from the above expressions that  $\chi^\pm(z)$ ($\Psi(z)$)  are two
different deformations
of a usual
real bosonic field (real fermionic field) and reduce to it in the limit
$q\rightarrow 1$.
The normal ordering symbol $:\cdots :$ introduced in \ref{ferm} means that the
annihilation
modes $\{a_{n\geq 0},b_{r\geq 1/2}\}$ are moved to the right of the creation
modes
 $\{a_{n<0},b_{r\leq -1/2}\}$ and the shift mode $a$.
In these, and in all other expressions in this paper,
operators and products of operators defined at the same point $z$
are understood to be normal ordered.
The relation \ref{reali} is the realization in terms of
one boson and
one fermion of $U_q(su(2)_2)$ that we will be using from now on.

 When $k=2$ the vertex operators can also be realized in terms of the same
fermionic field $\Psi(z)$ plus one deformed bosonic
field $\xi(z)$ given in terms of the same modes $\{a_n,a\}$ as
$\chi^\pm(z)$.
To be more specific, let $\xi(z)$ be
\be
\xi(z)=a-ia_0\ln{(- zq^4)}
+2i\sum\limits_{n>0}
{ q^{-3 n}a_n z^{-n}\over [2n]}
+2i \sum\limits_{n<0}
{q^{-5 n}a_n z^{-n}\over [2n]},
\ee
with
\be
\xi(z).\xi(w)=-\ln(q^4(wq^2-z))+:\xi(z).\xi(w):,\quad |z|>|q^2w|,
\ee
then the bare vertex operators
are realized as follows:
\be\br{rcl}
&\phi_1 (z)= \exp(i \xi (z)), \cr
 &\phi_0(z)=[\phi_1(z),E^{-}_0]_{q^2}=\sqrt{2}\oint
{ {d\omega} \ov {2 \pi i}} :\phi_1(z)\bE(\omega):
I_0(\omega,z) \psi(\omega),\\
&\phi_{-1}(z)=[\phi_0(z),E^{-}_0]/[2]=2\oint { {d\eta} \ov {2 \pi i}} \oint
{ {d\omega} \ov {2 \pi i}}
:\phi_1(z) \bE(\omega)\bE(\eta):\times \\ &I_0(\omega,z)
 (I^1_{-1} (\eta,\omega,z) \Psi(\omega).\Psi(\eta) +
I^{-1}_{-1} (\eta,\omega,z) \Psi(\eta).\Psi(\omega) ).
\label{BVO}\er\ee
Here
\be\br{rcl}
I_0(\omega,z)&=&
 { {(q^2 -q^{-2})} \ov {zq^2(1-q^6z/\omega)(1-q^{-2}w/z)} }, \quad
|zq^6|<|w|<|zq^2|,    \label{I_0}\\
I^1_{-1} (\eta,\omega,z)&=&{{ -(\omega-q^2\eta)} \ov{[2](zq^4) (1-q^{-2}
\eta/z)}},
\quad |q^{-2} \eta/z|<1 \label{I^1_{-1}},\\
I^{-1}_{-1}(\eta,\omega,z)&=& { {-(\eta-q^2\omega)} \ov {[2]\eta(1-q^6
z/\eta)}} ,
\quad |q^6z/\eta|< 1.\label{I^2_{-1}}
\er\ee
The denominators are left in the form $(1-a)$ in order that
the domain of convergence of the OPE sums $|a|<1$ is apparent. The $w$ and
$\eta$
integration contours must be chosen to lie within the domains indicated.
This point will be discussed more in Section 5.
Here the basic OPE's needed to derive (\ref{BVO})
are given by
\be\br{rcl}
\phi_1(z).\bE(\omega)=:\phi_1(z)\bE(\omega):{{-1}\ov{(zq^4)(1-q^{-2}\omega/z)}},
\\
\bE(\omega). \phi_1(z)= :\bE(\omega) \phi_1(z): { 1 \ov {\omega(1-q^6z/\omega)}
}\\
\bE(\omega_1).\bE(\omega_2)=:\bE(\omega_1)\bE(\omega_2):
(\omega_1-q^2\omega_2),
\er\ee
where $\bE(\omega)$ indicates the purely bosonic
part of $E^{-}(\omega)$.
In order to check our vertex operators \ref{BVO},  we have computed two matrix
elements involving  the vertex operators  \cite{BoWe93a}
\be
\vt^{\lambda_{2-i,V^1}}_{\lambda_{i}}(z)=(-zq^4)^{i/2}
(\phi_{+1}(z)\otimes v_{+1}^{1}+
\phi_0(z)\otimes v_0^1+
\phi_{-1}(z)\otimes v_{-1}^1),\qquad i=0,2.
\label{bvo}\ee
 After performing the various  double integrals, we
find
\be\br{rcl}
<\lambda_0|\vt^{\lambda_0,V^1}_{\lambda_2}(z)\vt^{\lambda_2,V^1}_{\lambda_0}
(w)|\lambda_0>&=&{1\over 1-q^4w/z}(q^2v_{+1}^1\otimes v_{-1}^1+v_{-1}^1
\otimes v_{+1}^1-q^2[2]
v_0^1\otimes v_0^1),\\
<\lambda_2|\vt^{\lambda_2,V^1}_{\lambda_0}(z)\vt^{\lambda_0,V^1}_{\lambda_2}
(w)|\lambda_2>&=&{1\over 1-q^4w/z}(v_{+1}^1\otimes v_
{-1}^1+q^2({w\over z})^2v_{-1}^1\otimes v_{+1}^1-q^2
[2]({w\over z})v_0^1\otimes v_0^1).
\label{matrixel}\er\ee
The above results coincide exactly with those given in formula 3.27
of Ref.  \cite{idzal93} which were
obtained by
solving the quantum Knizhnik-Zamolodchikov equation.

\subsection{Fock spaces\label{fock}}
{}From the realization \ref{reali} and \ref{BVO} it is clear that the
representation of the
$U_q(su(2)_2)$ are the Fock modules
\be\br{rcl}
F(n)&=&F_-|n>,\\
F_+|n>&=&0,
\er\ee
where $F_\mp$ are free ${\bf Q}(q)$ modules generated by $\{a_{\mp m}, b_{\mp
r},
m\in Z_{>0}, r\in {\bf Z}_{\geq 0}+1/2\}$. The states $|n>$, labelled by the
integers $n$, are defined by
\be
|n>=\exp(ina)|0>,
\ee
where $|0>$ is the `in' vacuum and it is annihilated by $\{a_m,m\geq 0;b_r,
 r\geq 1/2\}$. The dual Fock spaces are
defined
through $a\dag_n=a_{-n}$, $a\dag=a$, and $b\dag_r=b_{-r}$.
$F(n)$ is not
irreducible but splits into two irreducible Fock modules $F^0(n)$ and
$F^2(n)$, which are isomorphic   to
the representations $V(\lambda_0)$ and $V(\lambda_2)$ of
$U_q(su(2)_2)$ respectively. The latter
can be obtained from $F(n)$ through the following GSO (though not exactly
the usual GSO) projectors:
\be
P_\pm=(1\pm \exp{(-2\pi i d)})/2
\ee
as
\be\br{rcl}
F^0(n)&=&P_+F(n),\\
F^2(n)&=&P_-F(n),
\er\ee
and where
\be
-d=4\sl_{n>0}{n^2\over [2n]^2}a_{-n}a_n+2\sl_{r\geq 1/2}{r[2r]\over [4r]}
b_{-r}b_r+(a_0)^2/2.
\ee
Moreover, $F^0(n)$ and $F^2(n)$ are highest weight representations of
$U_q(su(2)_2)$, with highest weights $|0>$ and $|1>$ respectively.
The construction of the Fock module $F^{1}(n)$, which is isomorphic to the
representation $V(\lambda_1)$, is slightly more
complicated  \cite{Gin88}
 but as it is unnecessary for our purposes
we will not
discuss it here.

\section{Correlation Functions of Local Operators}

A `local operator' $L_N$ is one that acts naturally on the
$N$ tensor product $V\otimes V\otimes
\dots \otimes V$ ($V=V^{k/2}$) corresponding to $N$ adjacent sites of a
quantum spin chain. In order to interpret its action instead
in ${\cal F}=End(V(\lambda_i))$, which is the space of excitations
above the ground state labelled by $\lambda_i$,
it is necessary to introduce another set of type I
vertex operators
denoted by $\vt_{\sigma(\lambda_i),V}^{\lambda_i}(z)$
 \cite{FrRe92,Daval92,idzal93}. These vertex operators
intertwine
$U_q(su(2)_k)$ modules in the following way:
\be
\vt_{\sigma(\lambda_i),V}^{\lambda_i}(z):\quad V(\sigma(\lambda_i))\otimes
V(z)
\rightarrow V(\lambda_i).
\ee
Here $\lambda_{i}=(k-i)\Lambda_0+i\Lambda_1$ and
$\sigma(\lambda_i)=i\Lambda_0+(k-i)\Lambda_1$.
This VO has $k+1$ components $ \vt_{\sigma(\lambda_i),V,m}^{\lambda_i}(z)$
defined by
\be
\vt_{\sigma(\lambda_i),V,m}^{\lambda_i}(z)|u>=
\vt_{\sigma(\lambda_i),V}^{\lambda_i}(z)(|u>\otimes v_m)=
(id_{V(\lambda_i)}\otimes <v_m,.>)\vt^{\lambda_i, V^{*}}_{\sigma(\lambda_i)}
(z)|u>,
\ee
where $|u>\in V(\sigma(\lambda_i))$ and
\be
\vt^{\lambda_i, V^{*}}_{\sigma(\lambda_i)}(z)=
\alpha^{\lambda_i}_{\sigma(\lambda_i)}
(id_{V(\lambda_i)}\otimes C)\vt^{\lambda_i, V}_{\sigma(\lambda_i)}(zq^{-2}).
\ee
This implies that
\be
\vt_{\sigma(\lambda_i),V,m}^{\lambda_i}(z)=\alpha^{\lambda_i}_{\sigma(\lambda_i)}
{\cal C}_{-m}^{k/2}\vt^{\lambda_i, V}_{\sigma(\lambda_i),-m}(zq^{-2}),
\ee
where $\vt^{\lambda_i, V}_{\sigma(\lambda_i)}(z)=
\sl_{m=-k/2}^{k/2}\vt^{\lambda_i, V}_{\sigma(\lambda_i),m}(z)\otimes
v_m$ and $\alpha^{\lambda_i}_{\sigma(\lambda_i)}$ is a normalization constant,
which
we fix by \cite{idzal93}
\be
\vt_{\sigma(\lambda_i),V,k/2-i}^{\lambda_i}(z)|\sigma(\lambda_i)>=
\vt_{\sigma(\lambda_i),V}^{\lambda_i}(z)(|\sigma(\lambda_i)>\otimes v_{k/2-i})=
|\lambda_i>+\dots
\ee
Using in addition the normalization
\be
\vt^{\lambda_i, V}_{\sigma(\lambda_i),i-k/2}(z)|\sigma(\lambda_i)>=
|\lambda_i>+\dots
\ee
we arrive then at
\be
\alpha^{\lambda_i}_{\sigma(\lambda_i)}={1\over {\cal C}_{i-k/2}^{k/2}}.
\ee
This in turn  leads to
\be
\vt_{\sigma(\lambda_i),V,m}^{\lambda_i}(z)={{\cal C}_{-m}^{k/2}\over
{\cal C}_{i-k/2}^{k/2}}
\vt^{\lambda_i, V}_{\sigma(\lambda_i),-m}(zq^{-2}),\qquad -k/2\leq m\leq k/2.
\ee
If we set $k=2$ then $\sigma(\lambda_i)=\lambda_{2-i}$ and we get explicitly
\be\br{rcl}
\vt_{\lambda_i,V,\ep}^{\lambda_{2-i}}(z)&=&f_i(\ep)
\vt_{\lambda_i,-\ep}^{\lambda_{2-i},V}(zq^{-2}),
\label{vdown}\er\ee
where
$ f_i(+1) = q^{i-2}, \quad f_i(0)=-q^{i-2}/[2]$ and $f_i(-1)=q^i$.
Here $i=0,2$ because only the ground states labelled by $\lambda_0$ and
$\lambda_2$ \cite{idzal93} will be considered in this paper.
These ground states correspond to the two
antiferromagnetic spin configurations, $(\cdots +1,-1,+1,-1\cdots)$
and $(\cdots -1,+1,-1,+1\cdots)$ respectively. These are the
ground states for which the staggered polarisation for
the spin-1 XXZ model would be non-zero \cite{bax82,collin}.

{}From  the above definitions one can show  \cite{idzal93} that
\be
\vt_{\lambda_{2-i},V}^{\lambda_i}(z)\circ \vt^{\lambda_{2-i},
V}_{\lambda_i}(z)=
g_{\lambda_i}id_{V(\lambda_i)},
\ee
where $g_{\lambda_i}$ are scalar functions which can be
obtained from the two-point
matrix elements \ref{matrixel} and relation \ref{vdown}. We find then
\be
g_{\lambda_i}={1\over 1-q^2},\quad i=0,2.
\ee

The action of $L_N$ on ${\cal  F}$ can be specified by using the VOs
$\vt_{\l}^{\mu ,V}(z)$ and $\vt_{\l ,V}^{\mu}(z)$
 \cite{collin,idzal93}, in the
following way:
\bea \le{ {\cal L}_N^i= (g_{\l_i})^{-N} }\\
&&\!\!\!\!\vt_{\l_{i+2} ,V}^{\l_i}(z_1)\vt_{\l_{i+4} ,V}^{\l_{i+2}}(z_2)
\cdots \vt_{\l_{i+2N} ,V}^{\l_{i+2(N-1)}}(z_N)(id_{V_{\l_{i+2N}}}\otimes L_N)
\vt_{\l_{i+2(N-1)}}^{\l_{i+2N},V}(z_N)\cdots
 \vt_{\l_{i+2}}^{\l_{i+4} ,V}(z_2)\vt_{\l_i}^{\l_{i+2} ,V}(z_1) \nn\eea
Here $i=0$ or $2$ labels the choice of ground state
(or equivalently the $U_q(su(2)_2)$ module over which the trace is to be
taken), $z_1,\cdots,z_N$ are
local spectral parameters, and the subscripts on the $\l$s are understood as
modulo 4.
Then the correlation function of this operator
is given by:
\be <L>=L^{\epp_N\cdots \epp_1}_{\ep_N\cdots\ep_1}
P^{\epp_N\cdots \epp_1}_{\ep_N\cdots\ep_1}(z_1,z_2,\cdots,z_N|i),\label{<L>}\ee
where
\bea \le{P^{\epp_N\cdots \epp_1}_{\ep_N\cdots\ep_1}(z_1,z_2,\cdots,z_N|i) =
(1-q^2)^N \times } \\
&&Tr_{V(\l_i)}(q^{-2 \rho } \vt_{\l_{i+2} ,V,\epp_1}^{\l_i}(z_1)
\cdots \vt_{\l_{i+2N} ,V,\epp_N}^{\l_{i+2(N-1)}}(z_N)
\vt_{\l_{i+2(N-1)},\ep_N}^{\l_{i+2N},V}(z_N)\cdots
\vt_{\l_i,\ep_{1}}^{\l_{i+2} ,V}(z_1) )/Tr_{V(\l_i)}(q^{-2 \rho}), \nn\eea
and $\rho= 2d+a_0$ \cite{Daval92,collin,idzal93}.
 This trace may be rewritten purely in terms of the bare vertex operators of
\ref{BVO} as
\bea \le{P^{\epp_N\cdots \epp_1}_{\ep_N\cdots\ep_1}(z_1,z_2,\cdots,z_N|i) =
(1-q^2)^N\pl_{l=1}^N f_{(i+2l)_{ \bmod 4}} (\epp_l)
 (-z_l q^{4-(i+2l)_{\bmod 4}})
\times }\nn\\
&&\left(
T^{-\epp_1\cdots -\epp_N \ep_N\cdots\cdots\ep_1}(z_1q^{-2},\cdots,z_Nq^{-2},
z_N,\cdots,z_1|x,y)+\right.\nn \\
&&\left.(-1)^{i/2}T^{- \epp_1\cdots -\epp_N
\ep_N\cdots\cdots\ep_1}(z_1q^{-2},\cdots,z_Nq^{-2},
z_N,\cdots,z_1|{\bar x},y) \right)\times \nn\\
&&\left(Tr(x^{-d} y^{2a_0}) + (-1)^{i/2} Tr({\bar x}^{-d} y^{2a_0})
\right)^{-1}
\label{P}\eea
where $x=q^4,~{\bar x}^{1/2}=-x^{1/2},~y=q^{-1}$ and
\bea T^{\ep_1,\cdots,\ep_n}(z_1,\cdots,z_n|x,y)=
Tr(x^{-d} y^{2a_0} \phi_{\ep_1}(z_1) \cdots \phi_{\ep_n}(z_n)).
\label{trace}\eea
The traces in \ref{P} and \ref{trace} are taken over the complete Fock
space of Section \ref{fock}.
\section{Evaluation of the Trace}

The trace \ref{trace}
splits naturally into a product of contributions from the
bosonic non-zero modes, fermionic modes and bosonic zero modes \cite{ClSh73}.
The evaluation of these three terms will be dealt with separately.

\subsection{The trace over bosonic non-zero modes\label{bsect}}

{}From the expressions \ref{BVO} for $\phi_1(z), \phi_0(z)$ and $\phi_{-1}(z)$,
it is apparent that the complete
contribution of bosonic non-zero modes
to \ref{trace} is
\be Tr(x^{-d_b} \prod\limits_{i=1}^{n} F_{\epsilon_i}) \label{bt},\ee
where $-d_b=4\sl_{n>0}{ {n^2} \ov {[2n]^2}} a_{-n} a_{n}$, and
\be\br{rcl}
 F_{\ep_i=1}&=&\oP_1(z_i), \\
F_{\ep_i=0}&=&:\oE(\omega_i)\oP_1(z_i):, \\
F_{\ep_i=-1}&=&:\oE(\eta_i)\oE(\omega_i)\oP_1(z_i):.
\label{Fs}\er\ee
The hat indicates the bosonic non-zero mode part, and
$\omega_i$ and $\eta_i$ are integration variables.
To evaluate \ref{bt} we use the trace reduction technique of Clavelli
and Shapiro \cite{ClSh73}. The trace of an operator $O$ which is a function of
the
bosonic non-zero modes $a_n$, is given by
\be Tr (x^{-d_b} O) = { {<0|{\tilde O}|0>} \ov {(x;x)} },\ee
where $(s;x)\equiv \pl_{n=0}^{\infty}
(1-sx^n)$, and $\tilde O$ is the same operator expressed in terms of
oscillators ${\tilde a}_n$ defined by
\be {\tilde a}_n = { {a_n}\ov {1-x^n}} + c_{-n}~ (n>0), \quad
{\tilde a}_n = {a_n}+ {{c_{-n}} \ov {x^n-1}} ~(n<0).\ee
Here an  extra set of oscillators $c_n$ have been introduced which have the
same commutation relations as the $a_n$, and commute with them.
The technique for calculating $<0|{\tilde O}|0>$ is to completely normal
order the operator ${\tilde O}$  with respect to $a_n$ and $c_n$, and then to
use
Wick's theorem (for exponentials)
in order to express it in terms of contractions between
pairs of operators. When carrying out the integrations associated with
$\phi_0(z)$ and $\phi_{-1}(z)$, the contour must always be chosen
to be consistent with the domain of convergence of the series that
occur in the normal ordering calculation. Specifically this means
that whenever $1/(1 - a \omega) $ or $(aw;x)$
appears on the right-hand side of an OPE the contour for integration
over $\omega$ must be chosen
such that  $|a\omega |< 1$.

After normal ordering with respect to $a_n $ and $c_n$,
$\oP_1(z)$ and $\oE(\omega)$ become
\bea  \cP_1(z)& =& (q^2x;x) \exp\left( -2 \sum\limits_{n<0}
 { {(a_n q^{-5n} z^{-n} -c_n q^{3n} z^n)} \ov {[2n]}} \right)
\exp\left( -2 \sum\limits_{n>0}
{ {(a_n q^{-3n} z^{-n} -c_n q^{5n} (zx)^n)} \ov {(1-x^n)[2n]}} \right)\nn, \\
\cE(\omega) &=& (q^2x;x) \exp\left( 2 \sum\limits_{n<0}
{ {(a_n \omega^{-n} -c_n \omega^n) q^{-n}} \ov {[2n]} } \right)
\exp\left( 2 \sum\limits_{n>0}
{ {(a_n \omega^{-n} -c_n (\omega x)^n) q^n} \ov {(1-x^n)[2n]} }\right). \eea
Re-normal ordering with respect to $a_n$ and $c_n$ (indicated by $\no \cdots
\no$), gives
\bea
:\cE(\omega) \cP_1(z): &=&\no \cdots\no  { {(q^2x;x)^2} \ov {g(\omega/z) }}
\nn,\\
: \cE(\omega_1)\cE(\omega_2)\cP_1 (z): &=&\no \cdots \no
{ {(q^2x;x)^3 h(\omega_1/\omega_2)} \ov  {g(\omega_1/z)g(\omega_2/z)}},
\eea
where
\be\br{rcl} g(\omega/z)&=&(q^{-2}(\omega/z) x;x)
(q^6(z/\omega)x;x), \\
h(\omega_1/\omega_2)&=&(q^2 (\omega_1/\omega_2)x;x)
 (q^2(\omega_2/\omega_1) x;x).
\er\ee
Normal ordering in pairs gives
\be\br{rcl} \no\cP_1(z_1)\no\no \cP_1 (z_2)\no & =& \no\cdots \no G_1(z_1,z_2),
 \\
\no\cP_1(z)\no\no \cE(\omega)\no &=&\no\cdots\no G_2(z,\omega), \\
\no {\tilde E}^{-}(\omega)\no\no \cP_1 (z)\no  &=&\no \cdots \no G_3(\omega,z),
 \\
\no \cE(\omega_1)\no\no\cE(\omega_2)\no &=& \no\cdots\no
G_4(\omega_1,\omega_2),
\er\ee
where
\be\br{rcl}
G_1(z_1,z_2) & =&
(q^2(z_2/z_1);x)
(q^2(z_1/z_2) x;x),  \\
G_2(z,\omega)&=&
{1 \ov {(q^{-2}(\omega/z);x)(q^6(z/\omega)x;x)}},  \\
G_3(\omega,z)&=&
{ 1 \ov {(q^6(z/\omega);x) (q^{-2} (\omega/z)x;x)} },  \\
G_4(\omega_1,\omega_2)&=&
(q^2(\omega_2/\omega_1);x)
(q^2(\omega_1/\omega_2)x;x).
\er\ee
To label the fields we use the indices
\be\br{rcl}
a,\ap \in A &=&\{l|\ep_l=1,0,-1\},\quad {\rm dim}~A=n, \\
b, \bp \in  B &=&\{l|\ep_l=0,-1\},\quad {\rm dim}~B=n_B, \\
c, \cp \in C &=&\{l|\ep_l=-1\},\quad {\rm dim}~C=n_C.
\er\ee
Then in terms of these functions the non-zero bosonic mode contribution
\ref{bt} to the trace \ref{trace} is
\bea
 Tr(x^{-d_b} \prod\limits_{i=1}^{n} F_{\epsilon_i})&=&
{{(q^2x;x)^{n+n_B+n_C}} \ov {(x;x)}}
\pl_b {1 \ov {g(\omega_b/z_b)}} \pl_c {{h(\eta_c/\omega_c)}
\ov {g(\eta_c/z_c)}}\nn\\
&&\pl_{a<\ap}G_1(z_a,z_{\ap})\pl_{a<b}G_2(z_a,\omega_b)\pl_{a<c}G_2(z_a,\eta_c)
\\
&&\pl_{b<a}G_3(\omega_b,z_a)\pl_{b<\bp}G_4(\omega_b,\omega_{\bp})\pl_{b<c}
G_4(\omega_b,\eta_c)
\nn \\
&&\pl_{c<a}G_3(\eta_c,z_a)\pl_{c<b}G_4(\eta_c,\omega_b)
\pl_{c<\cp}G_4(\eta_c,\eta_{\cp}),
\nn \eea
where $\pl_b \equiv \pl_{b\in B}$ etc.
\subsection{The fermionic trace}
Introducing the set of integers $\{l_c\}$ with  $l_c=1$ or $-1$ for each
$c\in C$, which shall label both the order of the fermions and the associated
$I_{-1}^{l_c}(\eta_c,\omega_c,z_c)$ functions of \ref{BVO},
the fermionic contribution to \ref{trace} becomes
\be Tr(x^{-d_f} \pl_b H_b)\label{ft},\ee
where $-d_f=2 \sl_{r \ge 1/2} { {r [2r]}\ov {[4r]}} b_{-r} b_r $, and
\[ H_b= \left\{ \begin{array}{ll}
\Psi(\omega_b)\Psi(\eta_b),& b\in C,\qquad l_b=1,\\
\Psi(\eta_b)\Psi(\omega_b),& b\in C,\qquad l_b=-1, \\
\Psi(\omega_b),& b\not\in C.
\end{array}
\right.  \]
Using Clavelli and Shapiro's formula for fermionic trace reduction
 \cite{ClSh73} we get
\be Tr(x^{-d_f} O)=<0|{\tilde O}|0>\pl_{r \ge \half} (1+x^r).\ee
Now ${\tilde O}$ implies $b_r \ra {\tilde b}_r$, where,
\bea {\tilde b}_r = {{b_r} \ov {(1+x^r)}} + d_{-r}\quad (r\geq
\half)~, ~{\tilde b}_r =
b_r +{ {d_{-r}} \ov {(1+x^r)}}\quad (r\leq  -\half) .\eea
Here $d_r$ are anticommuting copies of the $b_r$.
Normal ordering with respect to $b_r$ and $d_r$ gives
\be {\tilde \Psi}(\omega_1).{\tilde\Psi}(\omega_2) =
\no\cdots\no ~+~ \D(\omega_1,\omega_2), \ee
where,
\be \D(\omega_1,\omega_2) = {{(\omega_1\omega_2)^{-\half} }
\ov 2} \sl_{r \geq \half} (q^{2r}+q^{-2r})
{ {((x\omega_1/\omega_2)^r +(\omega_2/\omega_1)^r)} \ov {(1+x^r)}} .\ee
This expression can be split into two parts, $\D_1(\omega_1,\omega_2)$ and
$\D_{-1}(\omega_1,\omega_2)$, where,
\bea \D_l(\omega_1,\omega_2)={{(\omega_1\omega_2)^{-\half} }\ov 2}
\sl_{r \ge \half}
{ {(q^{-2l} x\omega_1/\omega_2)^r +(q^{2l}\omega_2/\omega_1)^r} \ov {(1+x^r)}},
\quad l=\pm 1.
\eea
$\D_1(\omega_1,\omega_2)$, and $\D_{-1}(\omega_1,\omega_2)$ are convergent in
the strips $|x|<|q^2\omega_1/\omega_2|<1$ and
\newline
$|x|<|q^{-2}\omega_2/\omega_1|<1$ respectively. For the purpose of integration
$\D_l(\omega_1,\omega_2)$ may be re-expressed in terms
of the usual $\theta$ functions
through the identity of Goddard and Waltz \cite{GoWa71}:
\bea \D_l(\omega_1,\omega_2)={{i(\omega_1\omega_2)^{-\half} }\ov 4}
\theta_2(0|\tau)\theta_4(0|\tau)\theta_3(\nu_l|\tau)/\theta_1(\nu_l|\tau),\eea
where
$\nu_l=\ln(q^{2l}\omega_2/\omega_1)/(2\pi i)$ and
$\tau= \ln(x)/(2 \pi i)$.
Writing the theta functions in terms of infinite products  \cite{bateman}
gives
\be \D_l(\omega_1,\omega_2)={{(x^2;x^2)^2(x^{\half};x)^2} \ov
{2(q^{-l}\omega_1-q^{l}\omega_2)}}
{{(-x^{\half} (q^{2l}\omega_2/\omega_1);x)(-x^{\half}
(q^{2l}\omega_2/\omega_1)^{-1};x)}\ov {(x(q^{2l}\omega_2/\omega_1);x)
(x(q^{2l}\omega_2/\omega_1)^{-l};x)
}}.\ee
Then \ref{ft} is given by
\be Tr(x^{-d_f} \pl_b H_b) =(x^{\half};x) <0|\pl_b {\tilde H}_b|0> \ee
where the vacuum-vacuum correlation function is given by Wick's
theorem as a sum over all possible products of pair wise contractions
of the fermionic fields. Each of these contractions is given in terms
the fermionic ``propagator''
\be <\tilde{\Psi}(\omega_1)\tilde{\Psi}(\omega_2)> =
\D_1(\omega_1,\omega_2) + \D_{-1}(\omega_1,
\omega_2). \ee.
\subsection{The trace over zero modes}
The contribution of the bosonic zero modes to \ref{trace}
is the factor
\be Tr(x^{-d_0} y^{2a_0}\pl_{i=1}^n F^0_{\ep_i} ) \label{zt},\ee
where $-d_0=a_0^2/2$, and in analogy to equations \ref{Fs} we have
\be\br{rcl} F^0_{\ep_i=1}&=&\zP_1(z_i), \\
F^0_{\ep_i=0}&=&:\zE(\omega_i)\zP_1(z_i):, \\
F^0_{\ep_i=-1}&=&:\zE(\eta_i)\zE(\omega_i)\zP_1(z_i):,
\er\ee
with the zero sub/superscript indicating zero modes.
We define the functions
\be\br{rcl} \zP_1(z_1)\zP_2(z_2)=&:\zP_1(z_1)\zP_2(z_2):G^0_1(z_1,z_2),
\quad &G^0_1(z_1,z_2)=-zq^4,\\
\zP_1(z)\zE(\omega)=&:\zP_1(z)\zE(\omega):G^0_2(z,\omega),
\quad &G^0_2(z,\omega)=(-zq^4)^{-1},\\
\zE(\omega)\zP(z)=&:\zE(\omega)\zP(z):G^0_3(\omega,z),
\quad &G^0_3(\omega,z)=\omega^{-1},\\
\zE(\omega_1)\zE(\omega_2)=&:\zE(\omega_1)\zE(\omega_2):
G^0_4(\omega_1,\omega_2),
\quad &G^0_4(\omega_1,\omega_2)=\omega_1.
\er \ee
After completely normal ordering all operators in \ref{zt}, and provided
that $\sl_{i=1,n}\ep_i =0$, without which the trace vanishes,
the factor left
over is
\be Tr\left(x^{-d_0} y^{2a_0}\pl_a \exp(a_0 \ln(-z_aq^4)) \pl_b
\exp(-a_0\ln(\omega_b))
\pl_c \exp(-a_0 \ln(\eta_c))\right). \ee
Thus \ref{zt} is given by
\be\br{rcl}
Tr(x^{-d_0} y^{2a_0}\pl_{i=1}^n F^0_{\ep_i} )&=&
\sl_{m\in{\bf Z}} (x^{m^2/2} ( { {y^2{\bar z} (-q^4)^n } \ov { {\bar \omega}
{\bar \eta} }})^m ) \times  \\
&&\pl_{a<\ap}G^0_1(z_a,z_{\ap})\pl_{a<b}G^0_2(z_a,\omega_b)\pl_{a<c}G^0_2(z_a,
\eta_c)\\
&&\pl_{b<a}G^0_3(\omega_b,z_a)\pl_{b<\bp}G^0_4(\omega_b,\omega_{\bp})
\pl_{b<c}G^0_4(\omega_b,\eta_c)\\
&&\pl_{c<a}G^0_3(\eta_c,z_a)\pl_{c<b}G^0_4(\eta_c,\omega_b)
\pl_{c<\cp}G^0_4(\eta_c,\eta_{\cp}),
\er\ee
where ${\bar z}=\pl_a z_a,~{\bar \omega}=\pl_b {\omega_b}$ and ${\bar \eta}
=\pl_c {\eta_c}$.

\subsection{Collecting terms}
Defining ${\bar G_l}=G_lG^0_l,$ $(l=1,\cdots,4)$,  the complete expression
for \ref{trace} is
given by
\bea \le{T^{\ep_1,\cdots,\ep_n}(z_1,\cdots,z_n|x,y)=
(\sqrt{2})^{n_B+n_C} (q^2x;x)^{n+n_B+n_C}\times} \nn \\
&&\pl_b \oint {{d \omega_b} \ov {2 \pi i}}
I_0(\omega_b,z_b)
\sl_{\{l_c\}}
\left\{ \pl_c \oint {{d \eta_c} \ov {2 \pi i}}
 I^{l_c}_{-1}(\eta_c,\omega_c,z_c)
<0|\pl_b {\tilde H}_b|0>\right\} \nn \\
&&\pl_b {1 \ov {g(\omega_b/z_b)}} \pl_c {{h(\eta_c/\omega_c)} \ov
{g(\eta_c/z_c)}}\nn \\
&&\pl_{a<\ap}{\bar G}_1(z_a,z_{\ap})\pl_{a<b}{\bar G}_2(z_a,\omega_b)
\pl_{a<c}{\bar G}_2(z_a,\eta_c)\nn \\
&&\pl_{b<a}{\bar G}_3(\omega_b,z_a)\pl_{b<\bp}{\bar G}_4(\omega_b,
\omega_{\bp})\pl_{b<c}{\bar G}_4(\omega_b,\eta_c)\nn \\
&&\pl_{c<a}{\bar G}_3(\eta_c,z_a)\pl_{c<b}{\bar G}_4(\eta_c,\omega_b)
\pl_{c<\cp}{\bar G}_4(\eta_c,\eta_{\cp})\nn \\
&&\left[ {{(x^{\half};x)} \ov {(x;x)}}
\sl_{m\in{\bf Z}} \left(x^{m^2/2} \left( { {y^2{\bar z} (-q^4)^n}
 \ov { {\bar \omega}
{\bar \eta} }}\right)^m \right) \right]
\label{npoint},\eea
where the first sum is over all sets of $\{l_c\}$ with each $l_c=\pm 1$.
The contour for each integral in this sum is chosen in the way described
in section \ref{bsect}. Each contributing  integral  will correspond to
one choice
of $\{l_c\}$, and a particular fermionic contraction.
Using the Jacobi triple
product identity \cite{bateman},
\be  \sl_{m\in{\bf Z}} x^{m^2/2} z^{2m} = (x;x)(-x^{\half} z^2;x)
(-x^{\half} z^{-2};x), \ee
the last factor in \ref{npoint}
within $[\cdots ]$ can be written as
\be (x^{\half};x) (-x^{\half} ( { {y^2{\bar z} (-q^4)^n} \ov { {\bar \omega}
{\bar \eta} }});x) (-x^{\half} ( { {y^2{\bar z} (-q^4)^n} \ov { {\bar \omega}
{\bar \eta} }})^{-1};x). \ee
Similarly,
\bea Tr(x^{-d} y^{2a_0})&=&{(x^{\half};x) \ov (x;x)} \sl_{m \in {\bf Z}}
(x^{m^2/2}y^{2m})\nn\\
&=& (x^{\half};x) (-x^{\half}y^2;x)(-x^{\half}y^{-2};x). \eea
N-point correlation functions of local operators are then given in terms of
\ref{<L>}, \ref{P} and \ref{npoint}.

\section{Conclusions}

In this paper, we have derived an integral formula for
the N-point correlation functions of arbitrary local operators
of the antiferromagnetic spin-1
XXZ model in the thermodynamic limit. In doing this, we have derived a one
boson one fermion
realization of $U_q(su(2)_2)$.
This realization is more convenient than the q-deformation of the
Wakimoto realization of this algebra, which is given in terms of three deformed
bosons. This is because the latter realization
leads to infinitely many Fock spaces that are highly reducible.
To project out the irreducible subspaces, which are isomorphic to the highest
weight representations of $U_q(su(2)_2)$,  one has to study the cohomology
structure of the Fock spaces by means of the screening charges, which act
as BRST operators among them.
These screening charges are non-local and given as Jackson integrals in terms
of the screening currents. This means that had we used the q-deformation of
the Wakimoto construction, the N-point correlation functions would have
involved,
in addition to the usual integrals, both Jackson integrals
and
infinite summations arising from the BRST projections. These two would be
shortcomings are avoided in our one boson one fermion realization of
$U_q(su(2)_2)$ because the Fock spaces are already irreducible (up to simple
GSO-like projections as explained in section 3).

It is still a challenging
problem to try to  explicitly integrate, if not
the N-point correlations functions, at least the 1-point correlation functions.
Choosing the local operator as $S^z$
would lead, as for $k=1$ \cite{collin}, to the staggered polarisation
of the spin-1 XXZ model.
Other quantities of physical interest, which might be extracted  with this
technique,  are the local energy density, and form
factors of the model.

\section*{Acknowledgements}

We thank CRM for providing us with research fellowships and
a stimulating
environment. We wish to acknowledge and thank Amine El
Gradechi for many interesting discussions during the early
stages of this work. We are also grateful to many other members of CRM for
their
encouragement and advice, especially Luc Vinet and Yvan Saint-Aubin.
A.H.B. is highly appreciative to O. Foda for his encouragement and stimulating
communications. We thank Ian Affleck for pointing out reference
 \cite{what?} to us.

\baselineskip=10pt


\begin{thebibliography}{10}

\bibitem{Daval92}
B.~Davies, O.~Foda, M.~Jimbo, T.~Miwa, and A.~Nakayashiki.
\newblock {\em Comm. Math. Phys.}, 151:89, 1993.

\bibitem{tafa79}
L.A. Takhtadzhan and L.D. Faddeev.
\newblock {\em Russ. Math. Surveys}, 34:5:11--68, 1979.

\bibitem{KiRe87}
A.~N. Kirillov and N~Yu Reshetikhin.
\newblock {\em J. Phys.}, A20:1565, 1987.

\bibitem{Gaudin}
M.~Gaudin.
\newblock {\em La Fonction d'Onde de Bethe}.
\newblock Masson, 1983.

\bibitem{Aff89}
I.~Affleck.
\newblock {\em Field Theory Methods and Quantum Critical Phenomena}.
\newblock Les Houches, Session XLIX, 1988, {\it Champs, Cordes et
  Ph\'enom\`enes Critiques}, Ed. E. Br\'ezin and J. Zinn-Justin.

\bibitem{FrRe92}
I.~B. Frenkel and N.~Yu Reshetikhin.
\newblock {\em Comm. Math. Phys.}, 146:1, 1992.

\bibitem{collin}
M.~Jimbo, K.~Miki, T.~Miwa, and A.~Nakayashiki.
\newblock Correlation functions of the {X}{X}{Z} model for {${\D}<-1$}, 1992.
\newblock RIMS preprint.

\bibitem{bax82}
R.~J. Baxter.
\newblock {\em Exactly Solved Models in Statistical Mechanics}.
\newblock Academic, London, 1982.

\bibitem{ZaFa80}
A.B. Zamolodchikov and V.~A. Fateev.
\newblock {\em Sov. J. Nucl. Phys.}, 32(2):298, 1990.

\bibitem{KuRe81}
P.~P. Kulish and N.~Yu. Reshetikhin.
\newblock {\em Zap. Nauch. LOMI}, 101:101, 1981.

\bibitem{AlMa89}
F.C. Alcaraz and M.~J. Martins.
\newblock {\em J. Phys.}, A22:1829, 1989.

\bibitem{idzal93}
M.~Idzumi, T.~Tokihiro, K.~Iohara, M.~Jimbo, T.~Miwa, and T.~Nakashima.
\newblock {\em Int. J. Mod. Phys.}, A8:1479, 1993.

\bibitem{Sog84}
K.~Sogo.
\newblock {\em Phys. Lett.}, 104A:51, 1984.

\bibitem{what?}
M.~den Nijs and K.~Rommelse.
\newblock {\em Phys. Rev.}, B40:4709, 1988.

\bibitem{Res91}
N.~Reshetikhin.
\newblock {\em J. Phys, A: Math. Gen.}, 24:3299, 1991.

\bibitem{BoGr92}
A.H. Bougourzi and M.A.~El Gradechi.
\newblock Vertex realization of the ${U}_q(su(2)_2)$ quantum current algebra,
  1993.
\newblock {P}reprint {CRM}-1827, in press J. Group Theory Phys.

\bibitem{Abaal92}
A.~Abada, A.H. Bougourzi, and M.A.~El Gradechi.
\newblock Deformation of the wakimoto construction, 1992.
\newblock {P}reprint {CRM}-1829, in press Mod. Phys. Lett. A.

\bibitem{Mat921}
A.~Matsuo.
\newblock Free field representation of quantum affine algebra
  ${U}_q(\hat{sl}_2)$.
\newblock Nagoya University preprint, Aug. 1992.

\bibitem{Mat922}
A.~Matsuo.
\newblock Free field realization of q-deformed primary fields for
  ${U}_q({\widehat{sl}(2)})$, Dec 1992.
\newblock Nagoya University preprint.

\bibitem{Shi92}
J.~Shiraishi.
\newblock Free boson representation of ${U}_{q}(\hat{sl_{2}})$, 1992.
\newblock Preprint UT-617.

\bibitem{Katal92}
A~Kato, Y.-H. Quano, and J.~Shiraishi.
\newblock Free boson representations of q-vertex operators and their
  correlation functions, 1992.
\newblock Preprint UT-618.

\bibitem{Kim92}
K.~Kimura.
\newblock On free boson representations of the quantum affine algebra
  ${U}_q(\widehat{sl}_2)$.
\newblock 1992.
\newblock Kyoto University preprint.

\bibitem{FrJi88}
I.~B. Frenkel and N.~H. Jing.
\newblock {\em Proc. Nat'l. Acad. Sci. (USA)}, 85:9373, 1988.

\bibitem{Fel90}
D.~Bernard and G.~Felder.
\newblock {\em Comm. Math. Phys.}, 127:145, 1990.

\bibitem{Jim85}
M.~Jimbo.
\newblock {\em Lett. Math. Phys.}, 10:63, 1985.

\bibitem{Dri85}
V.~G. Drinfeld.
\newblock {\em Soviet Math. Doklady}, 32:254, 1985.

\bibitem{Dri88}
V.~G. Drinfeld.
\newblock {\em Soviet Math. Doklady}, 36:212, 1988.

\bibitem{ChPr91}
V.~Chari and A.~Presseley.
\newblock {\em Comm. Math. Phys.}, 142:261, 1991.

\bibitem{Ber89}
D.~Bernard.
\newblock {\em Lett. Math. Phys.}, 17:239, 1989.

\bibitem{bou93}
A.~H. Bougourzi.
\newblock Uniqueness of the bosonization of the ${U}_q(su(2)_k)$ quantum
  current algebra, 1993.
\newblock Preprint {CRM}-1852,in press Nucl. Phys. B.

\bibitem{sln}
H.~Awata, S.~Odake, and J.~Shiraishi.
\newblock Free boson realization of $u_q(\widehat{sl_N})$, 1993.
\newblock {RIMS-924}, {YITP/K-1018} preprint.

\bibitem{other}
H.~Awata, M.~Noumi, and S.~Odake.
\newblock Heisenberg realization for ${U}_q(sl(n))$ on the flag manifold, 93.
\newblock {YITP/K-1016} preprint.

\bibitem{vinet}
R.~Floreanini and L.~Vinet.
\newblock q-{D}ifference realizations of quantum algebras, June 93.
\newblock Preprint {CRM-1888}.

\bibitem{BoWe93a}
A.H. Bougouzi and R.~A. Weston.
\newblock Matrix elements of ${U}_q(su(2)_k)$ vertex operators via
  bosonization, 93.
\newblock Preprint {CRM-1875}.

\bibitem{Gin88}
P.~Ginsparg.
\newblock {\em Applied Conformal Field Theory}.
\newblock Les Houches, Session XLIX, 1988, {\it Champs, Cordes et
  Ph\'enom\`enes Critiques}, Ed. E. Br\'ezin and J. Zinn-Justin.

\bibitem{ClSh73}
L.~Clavelli and J.A. Shapiro.
\newblock {\em Nucl. Phys.}, B57:490, 1973.

\bibitem{GoWa71}
P.~Goddard and R.E. Waltz.
\newblock {\em Nucl. Phys.}, B34:99, 1971.

\bibitem{bateman}
Erd\'elyi et~al.
\newblock {\em Bateman Manuscript Project}, volume~2.
\newblock McGraw-Hill, 53.

\end{thebibliography}
\end{document}